# Parametric oscillation and amplification with gate controlled capacitor-within-capacitor

H. Grebel

Electronic Imaging Center at NJIT, Newark, NJ 07102. grebel@njit.edu

**Abstract:** A Capacitor-within-Capacitor (CWC) is a nested electronic element that has two components: the cell (e.g., the outer capacitor) and the gate (e.g., the inner capacitor). The designations for the gate and cell may be interchanged. Here, an analysis and experimentation on a diode-interfaced parametric oscillations and amplification are described. By replacing the diode with a junction, made of doped nano-graphene films, we demonstrated a new structure whose doping may be electronically and chemically controlled.

**I. Introduction:**

In equivalent circuit terms, capacitors may be connected in series (the overall capacitance becomes smaller than that of either capacitor), or in parallel (the overall capacitance is the sum of individual capacitances) [1]. Capacitors may take a simple form, such as two parallel plates, or a more complex structure of interdigitated electrodes [2]. Recently, a third possibility, a capacitor-within-capacitor (CWC), was considered [3]. Specifically, one capacitor, the gate capacitor, electronically controls the capacitance of a cell capacitor. The gate may be nested inside the cell (inner gate structure) or outside it (outer gate structure). Such a concept is general and may be applied to dielectric and super-capacitors alike. If the electrodes are made of 2-D films, such as graphene; voltage controlled charge doping of either electrode may be achieved in a rather simple manner [4].

Electrical parametric oscillators have been known for a long time [5]. In an electrical parametric oscillator, a resonating circuit is interfaced with a nonlinear capacitive, or nonlinear inductive [6] element. Modulation of the nonlinear element by a pump source at frequency $\omega_p$ and above some intensity threshold results in the generation of two frequency components: the signal at $\omega_s$ and an idler at $\omega_i$. Conservation of energy dictates that, $\omega_p = \omega_s + \omega_i$. When the signal and the pump are in phase (or, shifted by $\pi$ radians), then $\omega_s = \omega_i$ and the signal oscillates at half-frequency of the pump. Typically, a varactor, or a similar nonlinear capacitive element is used to realize the electrical circuit. Here, we take a somewhat different approach to realize a nonlinear capacitor (Fig. 1).

Consider an inner CWC structure (Fig. 1). A diode, placed across the gate capacitor [Ref. 3 and its SI section] provides for a voltage controlled element. From a circuit point of view, the effect of the gate diode on the cell's capacitance is understood as follows: if the diode is reversed biased (namely, the gate is open), then the structure is made of three capacitors connected in series. The cell's capacitance for this configuration is the smallest. If the diode is forward biased, then it shorts the gate and the cell's is made of only two capacitors in series; in this case the cell's capacitance increases. Alternatively, from an artificial dielectric point of view, one may consider the shortened gate as a giant electrical dipole whose increased polarity affects the cell's capacitance via an increase in the effective cell's permittivity. Similar arguments may be made for nonlinear artificial magnetic dipole, formed by the current loop that connects the pump source with the inner gate electrodes, the diode and the inductor and whose effect is to decrease the cell's permittivity value.

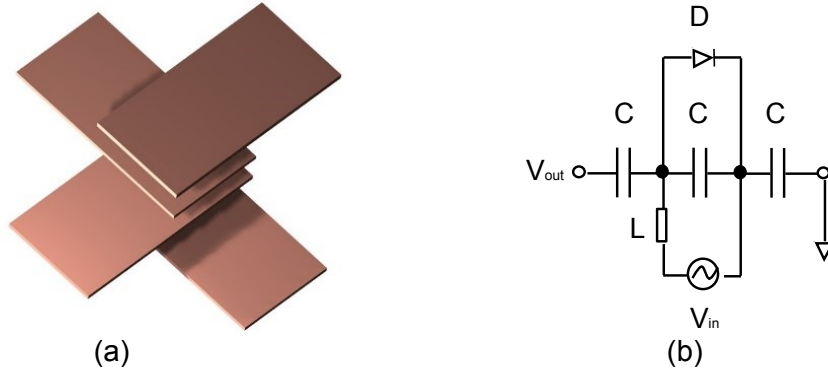

(a) (b)

Fig. 1. (a) Schematics of the CWC. Four stripes of copper are placed one on top of each other and are separated by dielectric films. (b) The equivalent circuit. C – capacitor; D – diode; L – inductor.

## II. Theoretical Considerations and Modeling:

Besides nonlinearities near the break-in region, a typical diode circuit exhibits nonlinearity when biased with either large negative amplitudes (the Zener effect), or large positive amplitudes (saturation effect). The effect at large positive biases happens when a small effective resistor, R, is connected in series with a diode. This resistor could be the result of contacts or wires. The current-voltage equation for the circuit current becomes, $I_d = I = I_0\{exp[(V-I \cdot R)/V_q]-1\}$, where $I_0$ is the dark current, V is the input voltage and $V_q$=26 mV at room temperature. For a large input voltage and small resistances, the term $I \cdot R$ competes with the input V and the current becomes saturated. Thus, the transition from a quasi linear I-V curve to saturated curve is also a useful parametric oscillation region.

Let us look closely at the nonlinear capacitive response. An equivalent circuit model is shown in Fig. 1b. One may assume that the capacitance, C, between nearest pairs of plates is the same because the spacing and the dielectric materials are similar. As described before, the maximum nonlinear swing may reach 50%, between cell capacitance values of C/3 to C/2 when the gate capacitor is either open or short.

We start with the nonlinear equation for a pumped capacitor [7]. We assume that the capacitance of the cell (namely, the outer set of electrodes) depends linearly on the shunt resistance across the gate within some resistance region [3]: $C(R_{shunt}) \sim C_0(1-R_{shunt}/R_0)$. Here $C_0$ is the cell capacitance without the shunt resistance and $R_0$ is a constant. From the diode equation for small currents (namely, ignoring the saturation region of the circuit), we can write for the effective diode resistance as a function of $V_p \equiv V_{pump}$,

$$R_{eff}=(dI/dV_p)^{-1}=(V_q/I_0)\{exp(-V_p/V_q)\}. \tag{1}$$

Combining, $C(V_p) \sim C_0(1-(V_q/R_0I_0)exp(-V_p/V_q))$. Finally, $\Delta C/C_0 \sim (V_q/R_0I_0)exp[-(V_{p0}/V_q)sin(\omega_p t)]$. As the voltage increases, the diode resistance decreases and the cell's capacitance increases.

The signal is set to the resonant frequency of this circuit, $\omega_s^2=(1/LC)$. We write, $C=C_0+\Delta C=C_0(1+\Delta C/C_0)$ and $L=L_0-\Delta L=L_0(1-\Delta L/L_0)$ with $\Delta C/C_0 \ll 1$ and $\Delta L/L_0 \ll 1$. Here $L_0$ and $C_0$ are associated with the resonance frequency of the gate. The nonlinear modulation of the gate affects the overall cell's capacitance and hence its frequency output. Note the choice of a plus sign to the capacitance change; the capacitance increases upon a positive pump swing.

The effect of L on the effective permittivity of the cell is expected to be smaller than the capacitive effect since magnetic dipoles are smaller than electric ones. In addition, the outer electrodes are well separated and form only a partial current loop. We will ignore it for the sake of simplicity and write for the parametrization of the frequency through first order expansion,

$$\omega_s^2 \approx (1/LC_0)(1-\Delta C/C_0)(1+\Delta L/L_0) \rightarrow (1/LC_0)(1-(V_q/R_0I_0)exp[-(V_{p0}/V_q)\sin(\omega_p t)]. \quad (2)$$

This is to be compared to a more traditional parametric oscillator,

$$\omega_s^2 \sim (1/L_0C_0)(1- bV_{p0}\sin(\omega_p t)), \quad (3)$$

where, b<<1 is a constant. Eq. (2) collapses to Eq. (3) for small signals. Pumping at twice the signal frequency, $\omega_p \sim 2\omega_{s0}=2\sqrt{(1/L_0C_0)} \sim 2\omega_s$ leads to a substantial signal gain and eventually to oscillations at the signal frequency.

Numerical results are shown in Fig. 2 for harmonic oscillator type differential equation. The parametrized radial frequency of either Eq. 2 or Eq. 3 was found by using a Mathcad tool. Specifically,

$$d^2y(t)/dt^2=-\omega_s^2 y(t)-\kappa dy/dt. \quad (4)$$

Here, the radial frequency is related to the frequency as, $\omega_s=2\pi f_s$ and $\kappa$ is the loss coefficient. In this semi ideal case, the loss coefficient in the simulations was small but not zero. Larger loss coefficients decrease the intensity of the frequency component. The various coefficients for the simulations were chosen such that Eq. 2 yields the same coefficients as Eq. 3 when expanding the exponent to first order of approximation. Specifically, when using Eq. 2: $(V_q/R_0I_0)=0.01$, $(V_{p0}/V_q)=2$, and when using Eq. 3: $bV_{p0}=0.02$. In the case of diode interfaced circuit, higher-order contributions give rise to a large gain in the signal. Finally, a fast Fourier transform (fft) module was used to assess the absolute value of the frequency components.

Boundary conditions for the simulations were: $y(0)=0$ and $(dy(t)/dt)_{t=0}=2\pi f_0=2\pi 0.55$. The normalized pump frequency varies, but it is approximately twice $f_0$, or, $f_p \sim 1.1$. With Eq. 2, the maximum gain is obtained with a somewhat lower or higher pump frequency than the pump frequency used with Eq. 3. Specifically, in the case of a parametric oscillator that is driven according to Eq. 3, the maximum gain to the Fourier component is found with $f_p=1.1$, for which $f_s=0.55$. In the case of a diode interfaced circuit, the maximum gain to the Fourier component is found with $f_p=1.085$ for which $f_s=0.542$ (Fig. 2). The experimental peak frequency also shifts as shown in Fig. 9 below.

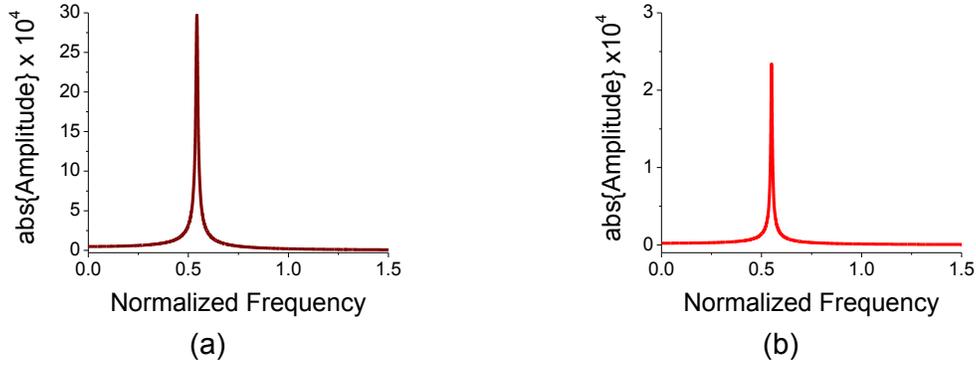

Fig. 2. Models: the absolute value of the Fourier component vs the normalized frequency, f. (a) When using Eq. 2: $(V_q/R_0I_0)=0.01$ and $(V_{p0}/V_q)=2$. (b) When using Eq. 3: $bV_{p0}=0.02$. The larger effect is observed for the highly nonlinear element - the diode (a).

In the simulations, a bandwidth of 0.4% of the central pump frequency, or Q~250 is noted. The effective parameters, chosen for the comparison between Eqs. 2 and 3 favor a relatively narrow band; these parameters are translated to a large effective circuit resistance, $R_0$.

Instead of looking at the leading Fourier component it might be instructive to watch the time evolution of the signal. A typical parametric oscillation, which is driven along with Eq. 3 would exhibit a monotonous, exponentially growing amplitude (Fig. 3b). In the case of a diode interfaced circuit, higher order nonlinearities affect the kinetics of the amplitude. The amplitude of the signal first decreases and then increases (Fig. 3a). Eventually, and as observed in the experiments, its signal grows much faster to provide a larger gain. Simulations also indicate that the amplitude minimum is shifting towards earlier times as the parameter $V_q/R_0I_0$ increases while keeping $V_{p0}/V_q$ constant. Similar trend is exhibited if we keep $V_q/R_0I_0$ constant while increasing $V_{p0}/V_q$; for example by increasing the pump amplitude, or by decreasing the temperature (and, thus decreasing $V_q=k_BT/q$, with $k_BT$ - the thermal energy and q - the electron charge).

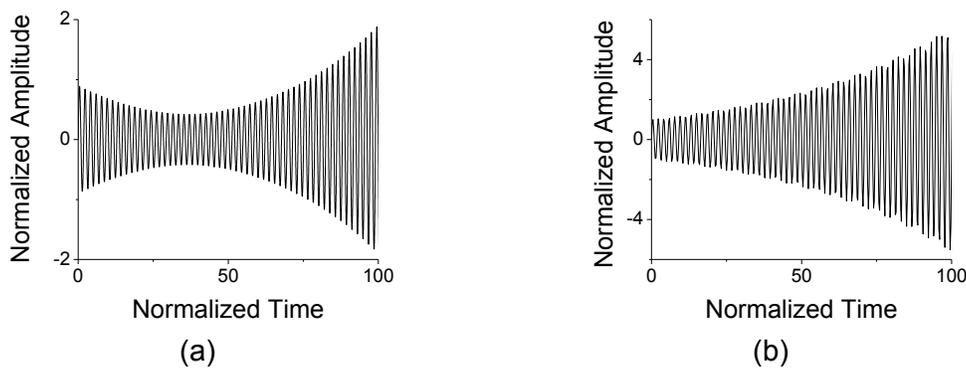

Fig. 3. Normalized amplitudes of the signal as a function of normalized time for: (a) parametric frequency of Eq. 2 and with parametric frequency of Eq. 3 (b). In the harmonic oscillator equation (Eq. 4), the Normalized Amplitude, y multiplies the square of normalized radial frequency of either Eq. 2, or Eq. 3. The normalized time is in units of $1/2\pi f_p$.

If the capacitance decreases upon a positive pump swing, we may choose a negative sign for the capacitance change. This is the case near the saturation region or when the diode direction is reversed. Hence, $C=C_0-\Delta C=C_0(1-\Delta C/C_0)$ and Eq. 2 now reads,

$$\omega_s^2 \approx (1/LC_0)(1+(V_q/R_0I_0)exp[-(V_{p0}/V_q)\sin(\omega_p t)]). \qquad (2')$$

For the simulations, parameters were selected as before: $(V_q/R_0I_0)=0.01$ and $(V_{p0}/V_q)=2$. Maximum gain is achieved with a slightly larger frequency than resonance, $f_p=1.115$, and the signal is exhibited at $f_s=0.558$. The Fourier component is now 5 time larger than the one depicted in Fig. 3a and the minimum in the temporal growth has shifted to t=0 (Fig. 4a).

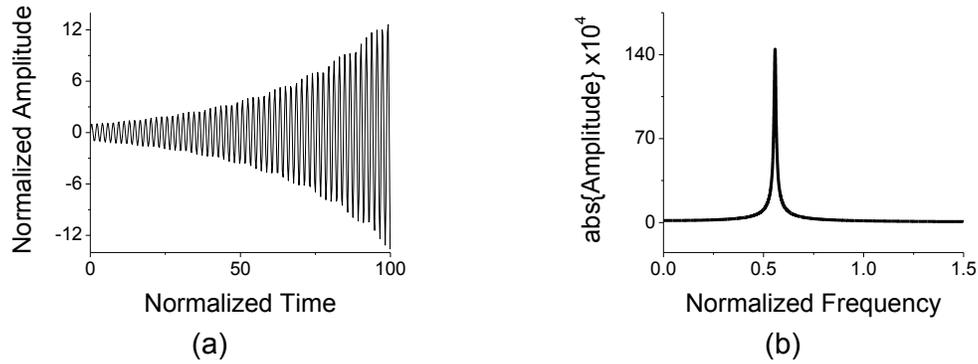

Fig. 4. Models: (a) Time evolution and (b) absolute value of the Fourier component vs the normalized frequency, f when using Eq. 2': $(V_q/R_0I_0)=0.01$ and $(V_{p0}/V_q)=2$.

## II. Methods and Experiments:

The CWC is made of 4 copper strips (plates), making an area of 12 mm x 12 mm when put across one another (Fig. 5a). They are separated by dielectric films (pieces of cut paper with slightly larger dimensions. The structure is held between two glass slides 25 mm x 25 mm by two clips. The inner gate, which controls the outer cell is operating at resonance with a quality factor of near unity. The capacitance is ca 0.014 nF and ca 0.04 nF between the outer plates and between the inner plates, respectively. The sinusoidal pump frequency is aimed at ca 1.06 MHz; the signal of this degenerate configuration is observed at ca 0.53 MHz (Fig. 6b). The 1N91 1DC723 Ge power diode that connects the gate electrodes has a large reverse breaking voltage rating ($V_R>75$ V) and a negative break-in voltage (Fig. 5b). Parametric oscillations can be observed with other Ge diodes, as well as with Si diodes (Fig. 5b), though at different DC bias; which is applied to gate electrodes. The entire circuit and layout are shown in Fig. 1c-d. A WaveTek frequency generator, an HP spectrum analyzer (SA) and HP oscilloscope are used to assess the input amplitude, the DC offset and the output spectrum.

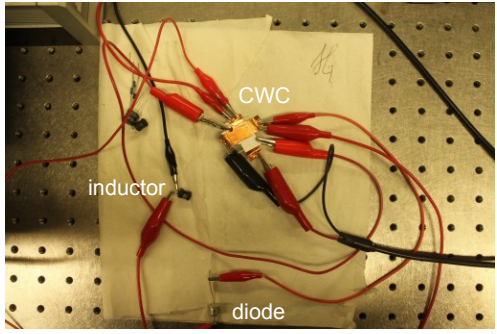 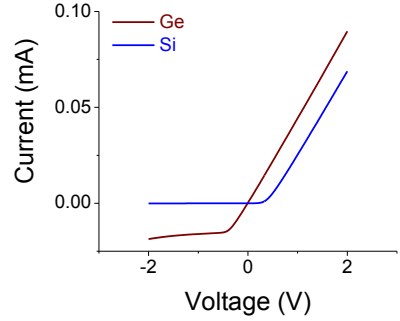

(a)                          (b)

Fig. 5. (a) A picture of the circuit. Four stripes of copper are placed one on top of each other and are separated by dielectric films. (b) I-V curve of Ge and Si diodes used in the experiments. A third Ge diode, whose I-V is similar to the Si diode, also enabled parametric oscillations. The output voltage Vout (Fig. 1b), is assessed with a spectrum analyzer (SA).

### III. Results and Discussions:

#### III.a. Experimental Results

As with other parametric oscillators, self-oscillation occurs beyond some pump threshold. In Fig. 6a we show the output signals just below the pump threshold; the pump amplitude is 1.8 $V_{p-p}$. The transmitted output is composed of only the pump frequency at 1.06 MHz. By increasing the pump amplitude to $V_{p-p}$=2 V and with a DC bias of ca 0.1 V (Fig. 6b), two additional peaks appear: at 0.53 MHz and at 1.59 MHz, respectively. The first peak is the parametric oscillation at half the pump frequency; the second peak is located at the pump frequency and the third one is the peak for the sum frequencies of pump and signal, $\omega_{3p/2} = \omega_p + \omega_s$. The input pump peak intensity was directly measured on the gate (including the inductor and the electrodes' capacitance), as -8.8 dBm for a 2 Vp-p amplitude. At threshold, the transmitted signal intensity at either $\omega_s$ or $\omega_{3p/2}$ is ca -40 dB of the direct input pump intensity, or -20 dB of the transmitted pump intensity. This is not an optimized configuration: one may achieve a transmitted signal which is only 3 dB below the transmitted pump. An intensity signal-to-noise ratio (SNR) of 40 dB is observed when increasing the input pump power by ca 1 dB from just below ($V_{pump}$=1.8 $V_{p-p}$) to just above threshold ($V_{pump}$=2 $V_{p-p}$). Note a decrease of 2.5 dB in the transmitted peak pump power due to generation of the sidebands (Fig. 6b in comparison to Fig. 6a). Fig. 6c shows that the oscillations are suppressed when reversing the diode connection while retaining the same pump conditions. This means that the polarity of the gate capacitor with respect to the polarity of the cell's capacitor, matters. The oscillations may be recovered with a proper negative or positive DC bias.

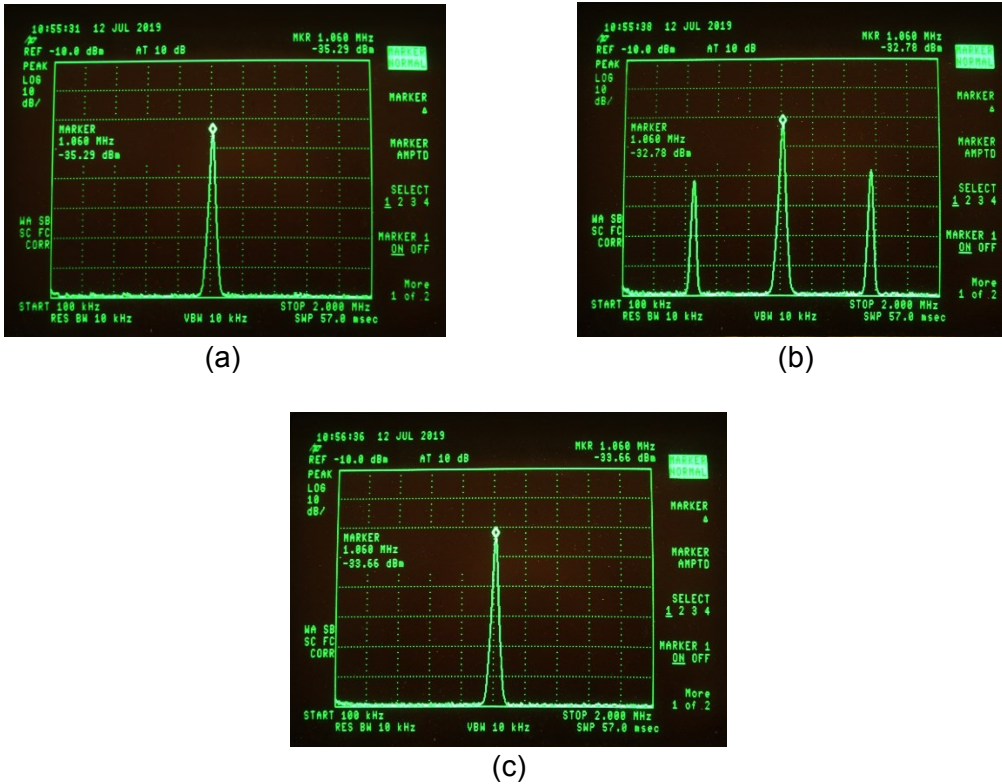

(a)　　　　　　　　　　　　　　(b)

(c)

Fig. 6. (a) Just below threshold. The marker is at 1.06 MHz. (b) Above threshold. Seen is the pump at 1.06 MHz, the signal at 0.53 MHz and the sum of the signal and the pump frequencies at 1.59 MHz. Also note that the transmitted pump power was reduced by 2.5 dB due to the generation of sidebands. (c) The diode connection is reversed while retaining the same conditions as in (b). The oscillation may be recovered by applying a proper DC bias.

Parametric oscillations rely on the nonlinearity of the diode resistance. At large DC bias, the diode exhibits a quasi linear I-V curve and its resistance is almost constant (Fig. 5b). Likewise, when reversed biased, the diode exhibits a constant (and a very large) resistance. These two regions are not appropriate for parametric oscillations. The diode's largest nonlinearity is near the break-in (forward bias), break down (reverse bias) and before the saturation (large forward bias) regions. For a sinusoidal pump amplitude of <2 $V_{p-p}$, the DC bias range for the Ge diode shown in Fig. 5b is between ±0.1 V and is optimal for 0 V.

Parametric oscillations may be observed for the above conditions when the pump frequency is scanned between f=0.45 to f=1.25 MHz. Within that bandwidth one may observe a rich spectra of sub-harmonics, high pump-frequency's harmonics, as well as combinations between them all (Fig. 7c). Based on the above bandwidth, the quality factor from the pump point of view, $Q=\omega_p/\Delta\omega=2\pi f_p/2\pi\Delta f$ is assessed as of order 1 (with $\omega_p$ chosen at the bandwidth center). The quality factor is also related to the resistance, capacitance and inductance as, $Q=R\sqrt{(C/L)}$. From the gate point of view with gate capacitance of ~0.04 nF and inductance of 0.22 mH we assess the resistance as, $R_{eff}$~2.3 KOhms. The impedance of the Ge diode in reverse DC bias is $R_r$=1.5 KOhms.

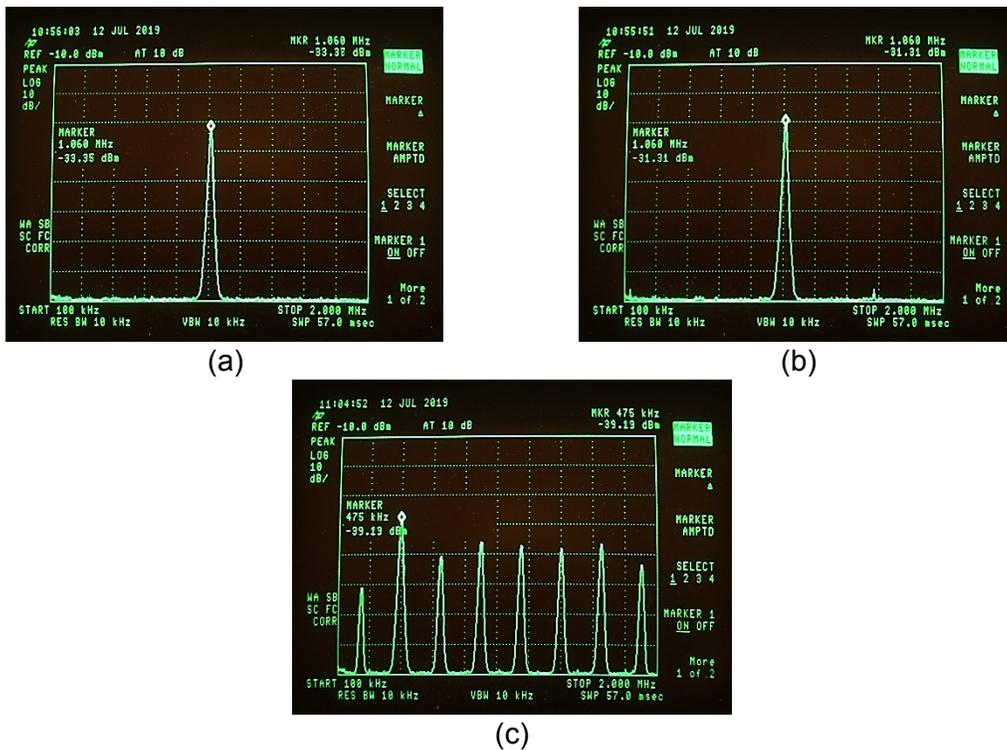

(a)                                 (b)

(c)

Fig. 7. (a,b) The DC offset is ±0.1 V and the oscillations are suppressed. (c) Full spectra with harmonics, sub-harmonics and related combinations when the pump frequency is at 0.475 MHz. Note that the pump power is smaller at this lower end of the bandwidth.

What happens if we increase the pump amplitude by 10 times its previous value, say to 20 $V_{p-p}$ or, 20 dB larger than the power threshold? Indeed, the transmitted pump intensity increases by 20 dB (from -30 dBm to -10 dBm); yet, the transmitted signal increases by 30 dB (from -50 dBm to -20 dBm (Fig. 8). The SNR of the signal is now 60 dB and its peak intensity is only 10 dB below the transmitted peak pump power. As mentioned before, the transmitted signal may reach -3dB of the transmitted pump at optimized conditions.

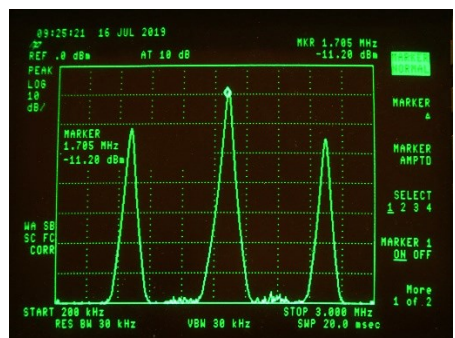

Fig. 8. Higher pump frequencies and broader signals are associated with a larger pump intensities. The signal-to-noise ratio (SNR) is 60 dB.

Fig. 9 shows a pump peak shift as the pump intensity increases.  Clearly, the pump frequency is up-shifting until the point of oscillation.  Similarly, upon changing the value of the inductor from L=0.22 mH to L=0.56 mH one may observe a pump frequency down shift (not shown).

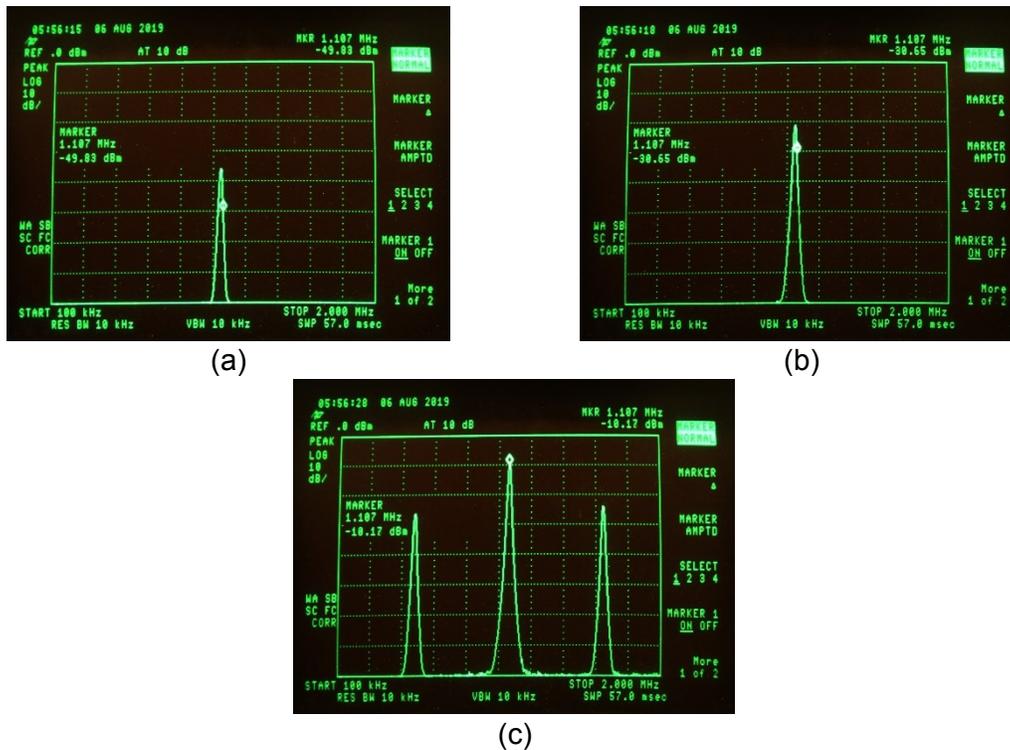

(a)

(b)

(c)

Fig. 9. (a-c) As the pump intensity increases throughout the 0-20 $V_{p-p}$, the pump frequency up-shifted

If one connects the first and third (negative) electrodes together for a shared ground keeping all other conditions the same, the resonance frequency shifts due to the 50 Ohms resistance of both the functional generator and the spectrum analyzer.  A useful frequency range was experimentally found at lower frequencies ($f_p$~600 KHz and $f_s$~300 KHz).

Graphene may be doped both chemically and electrostatically.  The recipe is based on similar approach used for single wall carbon nanotubes [8-9].  PVP-coated nano-graphene flakes exhibit p-type film and PEI coated graphene flakes exhibits n-type film.  Each of the film type was deposited on a 0.5-nm hole, TS80 membrane and the layout was pressed together. The flakes were soaked in the membrane, let dry out and contact to the film was made through the membrane back.  The thickness of the membrane reduced the capacitance of the gate, thus, the resonance of the CWC has increased to ca 3.1 MHz.  In Fig. 10 we present the related I-V curve, which exhibits a leaky characteristics.  Despite the small nonlinearity of the junction and without the Ge or Si diodes, oscillations were observed (Fig. 10b).  They are attributed to the further electrostatic doping of the graphene flakes.  The oscillations curve is much narrower than the curve obtained with a standard diode of Fig. 5.  Also, the oscillations frequency may be better tuned and do not necessarily occur at $f_{pump}/2$ but slightly above, or below it.

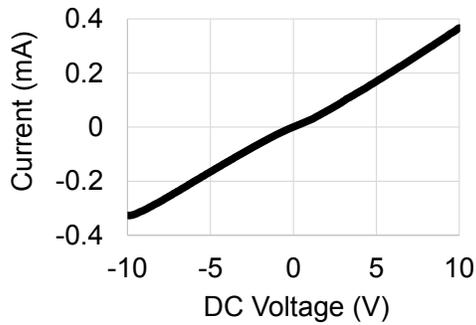
(a)
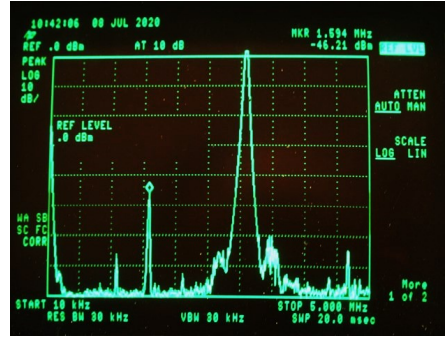
(b)

Fig. 10. (a) I-V curve of a junction made of chemically doped nano-graphene flakes. (b) Related oscillations.

### III.c. Amplification

These elements are efficient oscillators and exhibit amplification, as well. The circuit layout is shown in Fig. 11a. The transmitted pump intensity is -5 dBm, and the intensity of the nonamplified signalwas -50.44 dBm.  The pump power was adjusted to 1 dB below the onset of oscillations. As observed from Fig. 11a, the SNR of the signal was 30 dB and remained so upon amplification to -30.7 dBm (amplification of ca 20 dB).  Amplification of 30 dB was achievable if one adjusted the DC bias, as well as the pump power.  Finally, if the pump power was increased, oscillation mode took place.  The resistor R connected the grounds of the signal (Vs) and the pump (Vp) to the 50 Ohm ground of the spectrum analyzer (Vout).

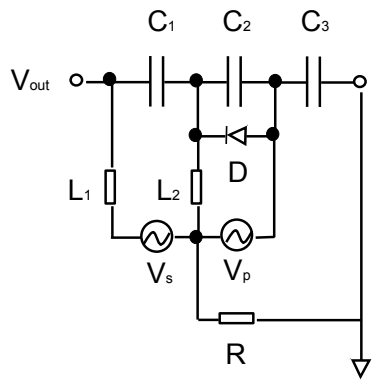
(a)

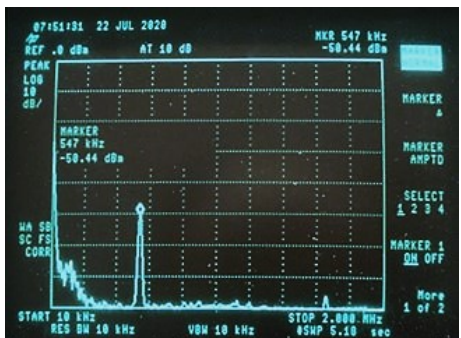
(b)

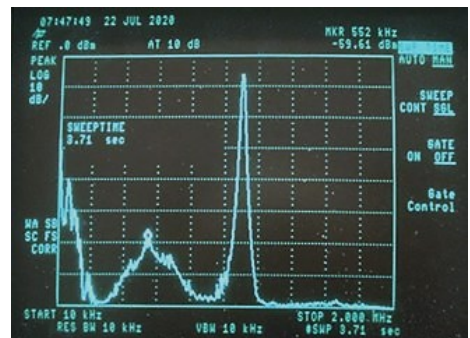
(c)

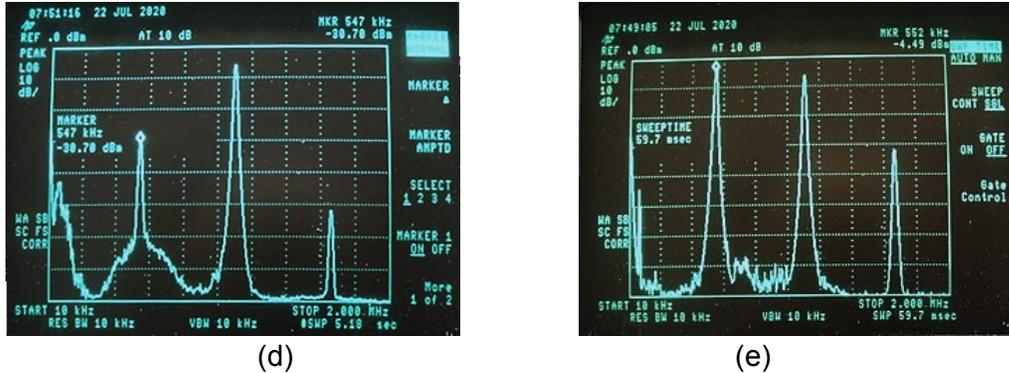

          (d)                                        (e)

Fig. 11. (a) The circuit layout. (b) The signal: -50.44 dBm at 547 KHz. (c) No-signal: the ~60 dBm amplifier noise just below the onset of oscillations. (d) The amplified signal at pump - 1 dB below the onset of oscillation and (e) oscillations at pump voltage of 20 $V_{p-p}$. $L_1=L_2=0.221$ mH and R=0.3 MOhms. Note that transmitted oscillations are above the transmitted pump.

## IV. Conclusions:

A simple and efficient parametric oscillator was built out of a nested structure, a capacitor-within-capacitor (CWC), a diode and an inductor. Higher orders and multiples of frequencies' sum and difference were exhibited due to large diode's nonlinearities [10]. A junction made of chemically doped graphene exhibited oscillations, as well, and may be useful for diode-like, or transistor-like structures, where a voltage bias shifts the Fermi level and hence the electrode's doping.

**Acknowledgement:**

1. L.D. Landau & E.M. Lifshitz, Electrodynamics of Continuous Media, (Volume 8 of A Course of Theoretical Physics), Pergamon Press 1960
2. Lei Zhu and Ke Wu, Accurate Circuit Model of Interdigital Capacitor and Its Application to Design of New Quasi-Lumped Miniaturized Filters with Suppression of Harmonic Resonance, IEEE Trans. Microwave Theory and Techniques, 48 (2000) 347–356.
3. Grebel, H., Capacitor within capacitor, SN Appl. Sci., 1 (2019) 48. doi.org/10.1007/s42452-018-0058-z
4. Inanc Meric, Melinda Y. Han, Andrea F. Young, Barbaros Ozyilmaz, Philip Kim, and Kenneth L. Shepard, Current saturation in zero-bandgap, top gated graphene field-effect transistors, Nature Nanotechnology, 3 (2008) 654-659.
5. Roer, T.G., Microwave Electronic Devices. Springer Science and Business Media, (2012) ISBN 978-1461525004.
6. K. D. Irwin and M. E. Huber, SQUID Operational Amplifier, Applied Superconductivity 2000, Sept. 17-22, 2000, Virginia Beach, Virginia.
7. A. Yariv, Quantum Electronics, 3$^{rd}$ edition, John Wiley and Sons, 1989.

8. T. Chowdhury and H. Grebel, Ion-Liquid Based Supercapacitors with Inner Gate Diode-Like Separators", ChemEngineering 2019, 3(2), 39; https://doi.org/10.3390/chemengineering3020039
9. Tazima S. Chowdhury, Haim Grebel, "Supercapacitors with electrical gates", Electrochimica Acta, Pub Date : 2019-04-02 , DOI: 10.1016/j.electacta.2019.03.222
10. Samuel Boutin, David M. Toyli, Aditya V. Venkatramani, Andrew W. Eddins, Irfan Siddiqi, and Alexandre Blais, "Effect of Higher-Order Nonlinearities on Amplification and Squeezing in Josephson Parametric Amplifiers", Phys. Rev. Appl. 8, (2017) 054030.